\documentclass[twocolumn,showpacs,preprintnumbers,amsmath,amssymb]{revtex4}

\usepackage{graphicx}
\usepackage{dcolumn}
\usepackage{bm}

\begin{document}

\preprint{}

\title{Effect of finite detection efficiency on the observation of the dipole-dipole interaction of
a few Rydberg atoms}
\author{I.~I.~Ryabtsev}
  \email{ryabtsev@isp.nsc.ru}
\author{D.~B.~Tretyakov}
\author{I.~I.~Beterov}
\author{V.~M.~Entin}
\affiliation{Institute of Semiconductor Physics\\ Prospekt Lavrentyeva 13, 630090 Novosibirsk, Russia }

\date{October 2, 2007}

\begin{abstract}
We have developed a simple analytical model describing multiatom signals that are measured in experiments on
dipole-dipole interaction at resonant collisions of a few Rydberg atoms. It has been shown that finite efficiency
of the selective field-ionization detector leads to the mixing up of the spectra of resonant collisions registered
for various numbers of Rydberg atoms. The formulas which help to estimate an appropriate mean Rydberg atom number
for a given detection efficiency are presented. We have found that a measurement of the relation between the
amplitudes of collisional resonances observed in the one- and two-atom signals provides a straightforward
determination of the absolute detection efficiency and mean Rydberg atom number. We also performed a testing
experiment on resonant collisions in a small excitation volume of a sodium atomic beam. The resonances observed
for 1-4 detected Rydberg atoms have been analyzed and compared with theory.
\end{abstract}

\pacs{34.10.+x, 34.60+z, 32.80.Rm, 32.70.Jz , 03.67.Lx}
 \maketitle

\section{INTRODUCTION}

Experimental and theoretical studies of long-range interactions of highly excited Rydberg atoms are important for
the development of quantum information processing with neutral atoms [1]. Strong dipole-dipole (DD) or van der
Waals interactions between Rydberg atoms allow for entanglement of neutral atoms and implementation of quantum
logic gates [2-7]. Practical realization of the proposed schemes for two-qubit phase gates [2,~4-6] and dipole
blockade [3,~7] demands the precise measurements of the number of Rydberg atoms, since two or single atoms must be
unambiguously detected in these schemes.

High detection efficiency of Rydberg atoms is thus a key issue for implementing the quantum information
processing. The most sensitive detection of Rydberg atoms is achieved with the selective field ionization (SFI)
technique that in addition allows for measurements of the population distribution [8]. A typical timing diagram of
the SFI detection is shown in Fig.1(a). A ramp of the electric field is applied after each laser excitation pulse
at delay $t_0$ that defines the time of free Rydberg-Rydberg interaction. Rydberg atoms ionize with almost 100\%
probability as soon as the electric field reaches a critical value $E_{cr} \sim n^{-4}$, where \textit{n} is the
principal quantum number. The resulting electrons or ions are detected either by a channeltron or by a
microchannel plate (MCP) detector.

For experiments with a few Rydberg atoms the channeltron is preferred, since the amplitudes of the channeltron's
output pulses are peaked near well-resolved equidistant maxima corresponding to the different numbers of detected
particles. As an example, in Fig.1(b) a histogram of the amplified output pulses of the channeltron VEU-6 (GRAN,
Russia) used in our experiments is presented. It exhibits the distinct maxima corresponding to the 1$-$5 electrons
from the sodium Rydberg atoms detected by SFI after the excitation by pulsed lasers.

Detection efficiency of channeltrons may reach 90\% [9]. However, several factors may significantly reduce the
overall SFI detection probability. These are metallic meshes of finite transparency $T=(50-90)\%$ used to form a
homogeneous electric field and to extract charged particles to the channeltron's input window [see Fig.4(a) for
our experimental arrangement], dependence of detection efficiency on the energy and type of charged particles
(electrons or ions), contamination of the channeltron's working surface after exposure to the atmosphere, etc.
These factors can negatively affect the signals and spectra measured in experiments on long-range interactions of
a few Rydberg atoms.

The main purpose of this paper is to analyze theoretically and experimentally the effect of reduced (and actually
unknown) detection efficiency on the observed spectra of resonant collisions of a few Rydberg atoms. Such
collisions are mediated by DD interaction and lie at the heart of the dipole-blockade effect and other related
schemes of quantum information processing. They were investigated in numerous experiments [10-24], but detection
statistics at resonance collisions of a few Rydberg atoms was not studied yet (although recently a sub-Poissonian
statistics of the MCP signals at the laser excitation of about 30 Rydberg atoms was studied both experimentally
[25] and theoretically [26,~27]). We have also implemented a new method to determine the absolute values of the
detection efficiency and mean number of Rydberg atoms excited per laser pulse. These issues are important for the
further development of quantum information processing with Rydberg atoms.

\begin{figure}
\includegraphics[scale=0.58]{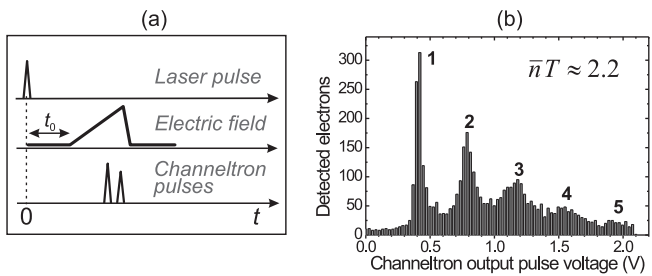}
\caption{\label{Fig1}(a) A typical timing diagram of experiments with selective field ionization (SFI) of Rydberg
atoms. (b) The histogram of the amplified output pulses of a channeltron used in our experiments. The observed
peaks correspond to 1$-$5 electrons from the sodium Rydberg atoms detected with SFI after the excitation by pulsed
lasers. The mean number of atoms detected per laser pulse  $\bar {n}T$ is about 2.2.}
\end{figure}

\section{Laser excitation and detection statistics for Rydberg atoms}

For \textit{N}$_{0}$ ground-state atoms in the laser excitation volume and probability of excitation for one atom
$0\le p \le 1$ the mean number of Rydberg atoms excited per laser pulse is

\begin{equation}
\label{eq1} \bar {n} = p\,N_{0}
\end{equation}

\noindent The statistics of the number of Rydberg atoms excited by each laser shot depends on \textit{p}. In the
case of \textit{weak} excitation (for example, by pulsed broadband laser radiation that is often used in
experiments) with $p \ll 1$ a Poisson distribution for the probability $P_{N}^{weak} $ to detect \textit{N}
Rydberg atoms after a single laser shot is applicable:

\begin{equation}
\label{eq2} P_{N}^{weak} = \frac{{\left( {\bar {n}} \right)^{N}}}{{N!}}\,\mathrm{e}^{ - \,\bar {n}}
\end{equation}

In the case of \textit{strong} excitation (for example, for coherent excitation of Rydberg atoms by narrow-band
lasers [28, 29]) a normal distribution must be used

\begin{equation}
\label{eq3} P_{N}^{strong} = p^{N}\left( {1 - p} \right)^{N_{0} - N}\frac{{N_{0} !}}{{N!\left( {N_{0} - N}
\right)!}}
\end{equation}

\noindent It is valid for any \textit{p} and \textit{N}$_{0}$ and provides general solutions for the measured
signals from Rydberg atoms, although analytical formulas obtained with Eq.(\ref{eq3}) may be rather complicated.

In the further analysis we will ignore a possible effect of the dipole blockade on the above distributions which
was discussed in Refs.~[25,~26], i.e., Rydberg-Rydberg interactions during exciting laser pulses will be
neglected. This is the case for appropriately short laser pulses. In this approximation, the above probability
distributions would be observed with an ideal SFI detector of Rydberg atoms. However, for a real detector with the
detection efficiency \textit{T} it can be shown that these distributions change to

\begin{equation}
\label{eq4} \bar {P}_{N}^{weak} = \frac{{\left( {\bar {n}T} \right)^{N}}}{{N!}}\,\mathrm{e}^{ - \,\bar {n}T}
\qquad \qquad \qquad \qquad \qquad
\end{equation}

\begin{equation}
\label{eq5} \bar {P}_{N}^{strong} = \left( {p\, T} \right)^{N}\left( {1 - p\, T} \right)^{N_{0} - N}\frac{{N_{0}
!}}{{N!\left( {N_{0} - N} \right)!}}
\end{equation}

The mean number of Rydberg atoms detected per laser shot thus reduces to $\bar {n}T$. This value can be measured
experimentally. For example, the amplitudes of the peaks in the histogram of Fig.1(b) are proportional to $\bar
{P}_{N}^{weak}$. Therefore, the relation between the integrated single-atom and two-atom peaks is $\bar
{P}_{2}^{weak} /\bar {P}_{1}^{weak} = \bar {n}T/2$, and our measurement gave $\bar {n}T \approx 2.2$. The
relations between the other integrated multiatom peaks are also well described by Eq.(\ref{eq4}) at $\bar {n}T
\approx 2.2$, confirming that the Poisson statistics is valid for the channeltron signals.

Thus, a pure interaction of, e.g., two Rydberg atoms cannot be observed with nonideal SFI detectors, since
measured two-atom signals would have a contribution from the larger numbers of Rydberg atoms. This is also true
for possible observations of the dipole blockade, which should appear as laser excitation of only one Rydberg atom
out of many interacting atoms due to the interaction-induced changes in the spectra of collective excitations [3].
Our aim is to analyze what detection efficiency is tolerable for experiments of such kind.

As the number of Rydberg atoms excited in each laser shot is unknown and fluctuates around $\bar {n}$, a
post-selection technique should be used in order to measure the signals for a definite number \textit{N} of
Rydberg atoms detected per laser pulse. In this technique the signals are first accumulated over many laser
pulses. Then they are sorted by the number of atoms  (up to 5 atoms in our experiments) according to the measured
in advance histogram of the channeltron's output pulses [Fig.1(b)]. After this procedure the signals are
separately determined for various \textit{N}. This technique has been demonstrated in our previous experiment on
microwave spectroscopy of multiatom excitations [5].

\section{Resonant collisions}

Dipole-dipole interaction of Rydberg atoms appears most prominently in resonant collisions (also called
F$\ddot{\mathrm{o}}$rster resonances), which have huge cross-sections [8]. Population transfer between Rydberg
states induced by such processes is a sensitive probe of DD interaction in atomic beams [10-13] and cold atom
clouds [14-24].

In the present work we have analyzed the Rydberg atom statistics and effect of finite detection efficiency on the
observed spectra of resonant collisions of a few sodium Rydberg atoms, both theoretically and experimentally. A
particular collisional resonance under study was the population transfer in the binary collisions:

\begin{equation}
\label{eq6} \mathrm{Na}(37\mathrm{S})+\mathrm{Na}(37\mathrm{S}) \to
\mathrm{Na}(36\mathrm{P})+\mathrm{Na}(37\mathrm{P})
\end{equation}

\noindent where Na(\textit{nL}) stands for a sodium Rydberg atom in highly-excited \textit{nL} state. Processes of
this kind were observed in many alkali-metal atoms and for various resonances. In the case of sodium, the energy
resonance arises when the $n$S state lies midway between the $n$P and $(n-1)$P states. The resonance of
Eq.(\ref{eq6}) is tuned by the Stark effect in the electric field 6.3-7.2 V/cm [Fig.2(a)], depending on the
fine-structure components of P-states. It appears as narrow peaks in the dependence of population of the 37P state
on the electric field after initial excitation of atoms to the 37S state. An example of collisional resonances
observed in our experiment with a velocity-selected atomic beam is shown in Fig.2(b) for the case of one Rydberg
atom detected with the post-selection technique. We note that the 37P sodium atoms are separately detected by SFI,
while the 37S and 36P atoms have almost identical critical electric fields for SFI and cannot be individually
detected.

\begin{figure}
\includegraphics[scale=0.51]{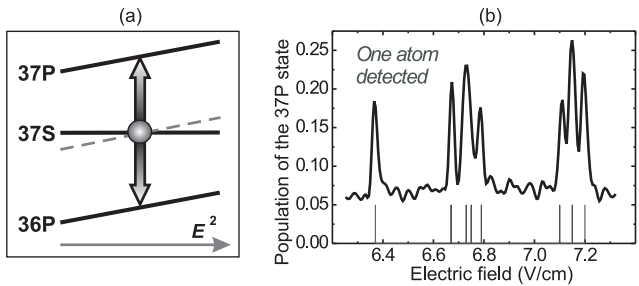}
\caption{\label{Fig2}(a) Sodium energy levels for the resonant collisions Na(37S)+Na(37S)$ \to $Na(36P)+Na(37P)
tuned by the electric field \textit{E}. (b) An example of the experimental spectrum of collisional resonances in a
velocity-selected atomic beam for the case of one sodium Rydberg atom detected. The vertical lines indicate the
calculated positions of resonances for various fine-structure components of P states.}
\end{figure}

Nonresonant background processes that can also populate the 37P state should be taken into account for the full
signal analysis. These are transitions induced by the ambient 300~K blackbody radiation (BBR) [8] and by
nonresonant collisions. They give a constant background signal, which is independent of the number of Rydberg
atoms and electric field [Fig.2(b)].

The signals of resonant collisions can be independently measured for various numbers \textit{N} of detected
Rydberg atoms by accumulating the statistics over many laser pulses and by their post-selecting. The normalized
\textit{N}-atom signals measured in our experiments are

\begin{equation}
\label{eq7} S_{N} = \frac{{n_{N} \left( {37\mathrm{P}} \right)}}{{n_{N} \left( {37\mathrm{S}} \right) + n_{N}
\left( {36\mathrm{P}} \right) + n_{N} \left( {37\mathrm{P}} \right)}}
\end{equation}

\noindent where $n_{N} \left( {nL} \right)$ is the total number of \textit{nL} Rydberg atoms detected by SFI
during the accumulation time for the particular case of \textit{N} detected Rydberg atoms. An example of the
experimental signal $S_{1} $ is shown in Fig.2(b).

For an ideal SFI detector Eq.(\ref{eq7}) simply gives the population $\rho _{N} $ (normalized per 1) of the 37P
state in a Rydberg atom after having interacted with (\textit{N}$-$1) surrounding Rydberg atoms. The approximate
formulas for $\rho _{N} $ will be obtained in Sec.V. For the nonideal detector, various $\rho _{N} $ contribute to
$S_{N} $ in a degree that depends on $\bar {n}T$, as it will be shown below.

\section{Multiatom signals at finite detection efficiency}

Expressions describing $n_{N} \left( {nL} \right)$ can be explicitly written for any \textit{N}:

\begin{equation}
\label{eq8} \left\{ {{\begin{array}{*{20}l}
 {n_{N} \left( {37\mathrm{P}} \right) =}  \\
 {\quad Z\sum\limits_{i = N}^{N_0}  {N P_{i} \,\rho _{i} \,T^{N}\left( {1 -
T} \right)^{i - N} \displaystyle\frac{{i!}}{{N! \left( {i - N} \right) !}}}}  \\
 {} \\
 {n_{N} \left( {37\mathrm{S}} \right) + n_{N} \left( {36\mathrm{P}} \right) =}  \\
 {\quad Z\sum\limits_{i = N}^{N_0}  {N P_{i} \,\left( {1 - \rho _{i}}
\right)\,T^{N}\left( {1 - T} \right)^{i - N} \displaystyle\frac{{i!}}{{N!\left( {i - N} \right)!}}}}  \\
\end{array}} } \right.
\end{equation}

\noindent where $Z \gg 1$ is the number of exciting laser pulses during the accumulation time and  $P_{i}$ is the
probability distribution given either by Eq.(\ref{eq2}) or Eq.(\ref{eq3}). Each term in the above sums represents
the contribution from \textit{i} actually excited Rydberg atoms, and it takes into account its statistical weight
and finite detection probability. Each $\rho _{i} $ should be viewed as a collision resonance contour, which would
be observed for the interaction of \textit{i} Rydberg atoms. The radiation dumping of Rydberg states is neglected
here, i.e., the interaction time is assumed to be much shorter than their effective lifetimes of $\simeq$30~$\mu
$s (37S state) and $\simeq$70~$\mu $s (36P and 37P states) at the 300~K ambient temperature.

Taking into account that

\begin{equation}
\label{eq9} n_{N} \left( {37\mathrm{P}} \right) + n_{N} \left( {37\mathrm{S}} \right) + n_{N} \left(
{36\mathrm{P}} \right) = Z N\bar{P}_{N}
\end{equation}

\noindent where $\bar {P}_N$  is the probability distribution given either by Eq.(\ref{eq4}) or Eq.(\ref{eq5}),
the measured, post-selected and averaged over $Z$ laser pulses signals of resonant collisions are

\begin{equation}
\label{eq10} S_{N}^{weak} = \mathrm{e}^{ - \bar {n}\left( {1 - T} \right)}\sum\limits_{i = N}^{\infty}  {\rho _{i}
\frac{{\left[ {\bar {n}\left( {1 - T} \right)} \right]^{i - N}}}{{\left( {i - N} \right)!}}}  \qquad
\end{equation}

\begin{eqnarray}
\label{eq11} S_{N}^{strong} = \displaystyle\frac{{\left( {N_{0} - N} \right)!}}{{\left( {N_{0} - \bar {n}T}
\right)^{N_{0} - N}}} \times \qquad \qquad \qquad \qquad & &\nonumber \\  \times \sum\limits_{i = N}^{N_{0}} {\rho
_{i} \displaystyle\frac{{\left[ {\bar {n}\left( {1 - T} \right)} \right]^{i - N}\left( {N_{0} - \bar {n}}
\right)^{N_{0} - i}}}{{\left( {N_{0} - i} \right)!\left( {i - N} \right)!}}}&&
\end{eqnarray}

\noindent For an ideal detector $(T=1)$ the expected identity $S_{N} = \rho _{N} $ is obtained with both
distributions, since only the $i=N$ term in the sum is nonzero.

We should note that $\rho _{1} $ cannot have a collision resonance feature, since this is a single-atom spectrum.
The population transfer in $\rho _{1} $ may appear only due to the mentioned background processes from BBR and
nonresonant collisions. At the same time, the other spectra (i.e., $\rho _{N > 1} $) must have a constant
non-resonant part identical to $\rho _{1} $ (as it is independent of the electric field) and a resonant part $\rho
_{N}^{res} $ from Rydberg-Rydberg interactions:

\begin{equation}
\label{eq12} \rho _{N > 1} = \rho _{1} + \rho _{N}^{res}
\end{equation}

The substitution of Eq.(\ref{eq12}) into Eq.(\ref{eq10}), Eq.(\ref{eq11}), and their reduction yield

\begin{equation}
\label{eq13} S_{N}^{weak} = \rho _{1} + \,\mathrm{e}^{ - \bar {n}\left( {1 - T} \right)}\sum\limits_{i =
N}^{\infty} {\rho _{i}^{res} \displaystyle\frac{{\left[ {\bar {n}\left( {1 - T} \right)} \right]^{i - N}}}{{\left(
{i - N} \right)!}}}
\end{equation}

\begin{eqnarray}
\label{eq14} S_{N}^{strong} = \rho_{1}+\displaystyle\frac{{\left( {N_{0} - N} \right)!}}{{\left( {N_{0} - \bar
{n}T} \right)^{N_{0} - N}}} \times \qquad \qquad \qquad & &\nonumber \\  \times \sum\limits_{i = N}^{N_{0}}
{\rho^{res}_{i} \displaystyle\frac{{\left[ {\bar {n}\left( {1 - T} \right)} \right]^{i - N}\left( {N_{0} - \bar
{n}} \right)^{N_{0} - i}}}{{\left( {N_{0} - i} \right)!\left( {i - N} \right)!}}}&&
\end{eqnarray}

\noindent In order to compare the observed amplitudes of collisional resonances, the nonresonant background $\rho
_{1} $ should be subtracted from these signals.

Equations (\ref{eq13}) and (\ref{eq14}) help to estimate the detection error and the contributions from the higher
numbers of actually excited Rydberg atoms. For example, the signal $S_{1} $ is important to study the dipole
blockade, where excitation of only one Rydberg atom by any laser excitation pulse is expected. With an ideal
detector ($T = 1$) the collisional resonance in $S_{1} $ must be absent. With nonideal detector ($T < 1$) the
resonance must also be absent at the full blockade, but it may appear at the partial blockade, when two or more
atoms are excited. Therefore, the amplitude of the resonance in $S_{1} $ can be a measure of the blockade
efficiency. Following Refs. [25,~26], the full blockade must also result in a complete disappearance of the peaks
with $N \ge 2$ in the histogram of Fig.1(b).

The signal $S_{2} $ is important to study two-atom interactions and to implement two-qubit quantum phase gate. We
see from Eq.(\ref{eq13}) that pure two-atom interactions in $S_{2} $ can be observed only when $\bar {n}\left( {1
- T} \right) \ll 1$, i.e., either $\bar {n}$ should be sufficiently small or $T$ must be close to 1. More
conclusions will be drawn after evaluating the dependences of $\rho _{N}^{res} $ on \textit{N} in the next
section.

\section{Evaluation of the multiatom spectra of resonant collisions}

Accurate calculation of $\rho _{N}^{res} $ is a challenging problem. It requires to account for many-body
interactions [30,~31], orientation of atom dipoles [18,~21,~32], time evolution of dipoles and populations
[23,~27,~31], spatial distribution of Rydberg atoms [18,~21,~33], coherent and pumping effects at the laser
excitation [20,~28,~29], etc.

On the other hand, as we are interested mainly in obtaining a scaling dependence of $\rho_N^{res}$  on $N$, this
problem can be significantly simplified by using the perturbation theory in a limit of weak DD interaction. For
this purpose we will introduce an average energy of DD interaction between \textit{any pair} of Rydberg atoms in
the atomic ensemble. This corresponds to the spatial and orientational averaging of laser excitation and of
Rydberg dipoles over the ensemble. Such approximation is reasonable, since we study the signals averaged over many
laser excitation and detection events in a disordered atomic ensemble.

The Hamiltonian of the ensemble is

\begin{equation}
\label{eq15} \hat {H} = \sum\limits_{k = 1}^{N_{0}}  {\hat {H}_{k} + \sum\limits_{n \ne m} {\hat {V}_{nm}} }
\end{equation}

\noindent where $\hat {H}_{k} $ is the unperturbed Hamiltonian of the \textit{k}th atom, and $\hat {V}_{nm} $ is
the operator of binary DD interaction of the arbitrary \textit{n}th and \textit{m}th atoms:

\begin{equation}
\label{eq16} \hat {V}_{nm} = \frac{{1}}{{4\pi \varepsilon _{0}} }\left[ {\frac{{\hat {\mathbf{d}}_{n} \hat
{\mathbf{d}}_{m}} }{{R_{nm}^{3}} } - \frac{{3\,\,\left( {\hat {\mathbf{d}}_{n} \mathbf{R}_{nm}}  \right)\,\left(
{\hat {\mathbf{d}}_{m} \mathbf{R}_{nm}} \right)}}{{R_{nm}^{5}} }} \right]
\end{equation}

\noindent Here $\hat {\mathbf{d}}_{n} $ and $\hat {\mathbf{d}}_{m} $ are dipole moment operators of the
\textit{n}th and \textit{m}th atoms, $\mathbf{R}_{nm} $ is a vector connecting these two atoms, and $\varepsilon
_{0}$ is the dielectric constant. The sum over $V_{nm}$ in Eq.(\ref{eq15}) accounts for all possible binary
interactions and contains $N(N-1)$ terms.

Our aim is to evaluate $\rho _{N}^{res} $ for a given number \textit{N} of actually excited Rydberg atoms. In
order to do that, we must decompose the wave function of the atom ensemble to a set of the collective states
corresponding to the different numbers of Rydberg atoms and then proceed with the calculations for certain
\textit{N}. Analytical calculations are possible only for a weak DD interaction, i.e., when the 37P state is
weakly populated. In the resonance approximation the two-atom transition amplitudes $a_{nm}$ for the process of
Eq.(\ref{eq6}) are found from

\begin{equation}
\label{eq17} a_{nm} \left( {t_0} \right) \approx - i \;\int\limits_{0}^{t_{0}} {\Omega _{nm} \left( {t}
\right)\,\mathrm{e}^{-i\Delta t}dt}
\end{equation}

\noindent Here $t_0$ is the interaction time and $\Delta = \left( {2\mathrm{E}_{37\mathrm{S}} -
\mathrm{E}_{36\mathrm{P}} - \mathrm{E}_{37\mathrm{P}}}  \right)/\hbar $ is the detuning from the resonance, which
is determined by the Stark-tuned energies E$_{nL}$ of the Rydberg states shown in Fig.2(a). The value $\hbar
\Omega _{nm}(t) $ is the energy of the resonant DD interaction between \textit{n}th and \textit{m}th atoms.

After calculating the time dependences of \textit{a}$_{nm}$, the desired spectra $\rho _{N}^{res} $ can be found
from

\begin{equation}
\label{eq18} \rho _{N}^{res} = \frac{{1}}{{N}}\sum\limits_{n \ne m} {\left| {a_{nm}} \right|^{2}}
\end{equation}

\noindent Each value $\left| {a_{nm}}  \right|^{2}$ is the probability of two atoms $n$ and $m$ out of \textit{N}
Rydberg atoms to simultaneously leave the initial 37S state and undergo the transitions to the final 36P and 37P
states. The factor 1/\textit{N} appears because in this final collective state the population of the 37P state per
atom is $1/N$. Other collective states with two or more 37P atoms are not populated at weak DD interaction.
Therefore, the main problem is the calculation of $a_{nm}$ and $\rho_N^{res}$ for particular experimental
conditions. We will consider a frozen Rydberg gas and an atomic beam, which differ in the time dependence of the
interaction energy.

\subsection{Frozen Rydberg gas}

In a frozen Rydberg gas [14,~15] the atoms are almost immobile during the free interaction time [$t_0$  in
Fig.1(a)] of a few microseconds [23]. In this case $\Omega _{nm} $  are constants, and $a_{nm}$ are readily
calculated from Eq.(\ref{eq17}). Then Eq.(\ref{eq18}) yields

\begin{equation}
\label{eq19} \rho _{N}^{res} \left( {t_{0}}  \right) \approx \frac{{\sin^{2\,}\left( {\,t_{0} \Delta /2}
\right)}}{{N\;\left( {\Delta /2} \right)^{2}}}\sum\limits_{n \ne m} {\left| {\Omega _{nm}}  \right|}^{2}
\end{equation}

The full width at half maximum (FWHM) of this resonance is $ \approx 1/t_{0} $ and it is independent of \textit{N}
at the weak DD interaction. For the analysis of its amplitude we may note that the sum in Eq.(\ref{eq19}) has
$N(N-1)$ terms and therefore can be represented as

\begin{equation}
\label{eq20} \sum\limits_{n \ne m} {\left| {\Omega _{nm}}  \right|^{2} \equiv N\left( {N - 1} \right)\,\Omega
^{2}}
\end{equation}

\noindent where $\hbar \Omega $ is the mean-square energy of resonant DD interaction between \textit{any pair} of
atoms in the ensemble. The averaging must be made over all possible positions of 2 Rydberg atoms among $(N_{0}-2)$
ground-state atoms in the laser excitation volume. We note that Eq.(\ref{eq20}) has a general form that gives a
correct result even at $N=1$, when DD interaction must be absent.

The value of $\Omega^2$  depends on the atom density and geometry of the excitation volume [21]. Its accurate
analytical or numerical calculation must be a subject of a special study and lies beyond the scope of this paper.
We shall only estimate it in the following qualitative manner.

Let us consider the DD interaction between two arbitrary Rydberg atoms in the laser excitation volume containing
$N_{0} \gg 1$  ground-state atoms. A peculiarity of this problem is that after the laser pulse any atom can be
found in a Rydberg state. The number density $n_0$ of ground-state atoms defines the average spacing $R_{0}
\approx \left( {4\pi n_0 /3} \right)^{ - 1/3}$ between two neighboring atoms. The energy of DD interaction of two
Rydberg atoms spaced by $R_0$ is estimated from Eq.(\ref{eq16}) as

\begin{equation}
\label{eq21} \hbar \Omega _{0} \sim \frac{\left\langle {36\mathrm{P}\left| {\hat {\mathbf{\mathrm{d}}}}
\right|37\mathrm{S}} \right\rangle \left\langle {37\mathrm{P}\left| {\hat {\mathbf{\mathrm{d}}}}
\right|37\mathrm{S}} \right\rangle }{{4\pi \varepsilon _{0} R_{0}^{3}} }
\end{equation}

Our aim is to express $\Omega^2$ through $\Omega _{0} $ and $N_{0} $. In the averaging procedure for $\Omega^2$ we
will use the below qualitative argumentation that seems to be valid for disordered atom ensembles at $N_0 \gg N$.
Let the excitation volume has a spherical shape of radius $R\gg R_{0} $, so that the volume is $\mathrm{V} = 4\pi
R^{3}/3$. We have to find a mean-square energy of DD interaction of an atom, situated deeply in this volume, with
the surrounding $(N_{0}-1) \approx N_{0}$ atoms. The mean atom number in the volume of radius $R_{0} $ is $4\pi
R_{0}^{3} n_{0} /3 = 1$. This implies that on the average there are no other atoms at the distances $r < R_{0} $
from this central atom. Therefore, one should only account for the interactions of this atom with other atoms
situated at $r \ge R_{0} $. The mean number of atoms at the distances lying in the interval from $r$ to $r + dr$
is $\left( {n_{0} 4\pi r^{2}dr} \right)$. Hence, the statistical weight of this spherical layer is $\left( {n_{0}
4\pi r^{2}dr/N_{0}} \right)$. Then the averaging is done with the integral

\begin{equation}
\label{eq22} \Omega ^{2} = \frac{{n_{0}} }{{N_{0}} }\int\limits_{R_{0}} ^{R} {\frac{{\Omega _{0}^{2} R_{0}^{6}}
}{{r^{6}}}\;4\pi r^{2}dr \approx \frac{{\Omega _{0}^{2}} }{{N_{0}} }} \;
\end{equation}

\noindent The relations $N_{0} = n_{0} \mathrm{V}$ and $R \gg R_{0} $ have been applied here. The physical meaning
of Eq.(\ref{eq26}) is that the main contribution to $\Omega ^{2}$ comes from the nearest-neighbor atom at $r
\approx R_{0} $, which can be found in a Rydberg state with the $1/N_{0} $ probability. Other atoms on the average
are situated at larger distances and their contribution has been estimated to be small. Although Eq.(\ref{eq26})
gives only a rough estimate for the mean-square energy of DD interaction of two arbitrary Rydberg atoms in large
disordered atom ensembles, it can be used in the approximate analytical calculations instead of more precise but
complicated numerical simulations that must account for the actual geometry of the excitation volume.

Finally, we find that for the weak DD interaction the amplitude of the resonance scales as

\begin{equation}
\label{eq23} \;\rho _{N}^{res} \left( {t_{0} ,\;\Delta = 0} \right) \sim (N-1)\; \frac{{\Omega _{0}^{2}} }{{N_{0}}
} \;t_{0}^{2}
\end{equation}

\subsection{Atomic beam}

Atoms in thermal atomic beams have a wide velocity spread (on the order of $v_{0} = \sqrt {2k_{\mathrm{B}}
\mathrm{T}/M} \approx 680$~m/s at the $\mathrm{T}=650$~K temperature for Na, where $k_{\mathrm{B}} $ is the
Boltzmann constant and \textit{M} is the atom mass). Therefore, the energy of DD interaction of a pair of Rydberg
atoms is most likely a rapidly varying function of time [10-13]:

\begin{equation}
\label{eq24} \Omega _{nm} \left( {t} \right) \sim \Omega _{0} \frac{{R_{0}^{3}} }{{\left[ {v_{nm}^{2} \left( {t -
t_{nm}}  \right)^{2} + b_{nm}^{2}}  \right]^{3/2}}}
\end{equation}

\noindent where $v_{nm} $ is the relative velocity of the two atoms, $b_{nm} $ is the impact parameter of
collision, and $t_{nm} $ is the time moment when the interaction energy reaches its maximum. In the case of atomic
beam both energy and time of interaction depend on the initial distance between the two Rydberg atoms.

At $v_{nm} \gg \left( {b_{nm} /t_{0}}  \right)$ each pair of Rydberg atoms interacts momentarily and only one
time. The effective collision time obtained from Eq.(\ref{eq24}) is $\tau _{nm} \approx 2b_{nm} /v_{nm} $. By
analogy with Eq.(\ref{eq23}), the estimated amplitude of the resonance for \textit{N} Rydberg atoms at the weak DD
interaction is

\begin{equation}
\label{eq25} \;\rho _{N}^{res} \left( {t_{0} ,\;\Delta = 0} \right) \sim (N-1)\; \Omega ^{2}_{beam}\tau^{2}
\end{equation}

\noindent where $\tau $ is the mean collision time and $\hbar \Omega_{beam} $ is the mean-square interaction
energy for any of the two atoms in the atomic beam. We may expect that the main contribution to the transition
amplitude is also from the nearest-neighbor atoms. Then the collision time is $\tau \sim 2R_{0} /v_{0} $ and the
width of the resonance is $\Gamma \sim 1/\tau $.

The value of $\Omega^2_{beam} $ differs from $\Omega^2 $ in Eq.(\ref{eq22}). Due to the relative motion of the
atoms in the beam, each atom effectively interacts with the larger amount of atoms, than in the frozen gas. The
approximate formula $\Omega _{beam}^{2} \sim\Omega _{0}^{2} /N_{0}^{2/3} $ has been obtained from averaging of
Eq.(\ref{eq24}) over the impact parameter instead of the interatomic distance in Eq.(\ref{eq22}). A more accurate
calculation of the resonance line shape for the atomic beam must be a subject of a special study. We only note
that averaging over the velocity distribution may result in the cusp-shaped resonances (pointed at the top) due to
the longer interaction time of atoms with low collision velocities [13].

\subsection{Multiatom signals at weak DD interaction}

Multiatom signals of Eq.(\ref{eq13}) and Eq.(\ref{eq14}) can be further reduced using Eq.(\ref{eq23}) or
Eq.(\ref{eq25}). The explicit \textit{N} dependences of these equations allow us to represent any $\rho _{N}^{res}
$ through the two-atom spectrum $\rho _{2}^{res} $ as

\begin{equation}
\label{eq26} \rho _{N}^{res} = \left( {N - 1} \right)\,\rho _{2}^{res}
\end{equation}

The substitution of Eq.(\ref{eq26}) into Eq.(\ref{eq13}) and Eq.(\ref{eq14}) gives after the sum reduction:

\begin{equation}
\label{eq27} S_{N}^{weak} = \rho _{1} + \rho _{2}^{\mathrm{res}} \left[ {N - 1 + \;\bar {n}\left( {1 - T} \right)}
\right]
\end{equation}

\begin{equation}
\label{eq28} S_{N}^{strong} = \rho _{1} + \rho _{2}^{\mathrm{\mathrm{res}}} \left[ {N - 1 + \;\bar {n}\left( {1 -
T} \right)\frac{{1 - \bar {n}/N_{0}} }{{1 - \bar {n}T/N_{0} }}} \right]
\end{equation}

\bigskip
It is seen that Eq.(\ref{eq27}) can be obtained from Eq.(\ref{eq28}) at $N_{0} \gg \mathrm{max}\left( {1,\;\bar
{n}T} \right)$. This is a criterion for the validity of the Poisson distribution, which means that the number of
all atoms in the excitation volume must be much larger than 1 or than the number of detected Rydberg atoms.

The further analysis of signals will be performed with Eq.(\ref{eq27}), which is applicable to most of the
experimental conditions. For the correctness of measurements, the resonances in $S_{1} $ must be small compared to
those in $S_{N > 1} $. This is the case if $\bar {n} \ll 1/\left( {1 - T} \right)$, i.e., if the mean Rydberg atom
number is limited to some maximum allowed value. This value is large if only $T$ is close to 1.

At $N>1$ the criterion of correct measurements changes to $\bar {n}\ll \left( {N - 1} \right)/\left( {1 - T}
\right) $. In this case $S_{N} \approx \rho _{1} + \rho _{N}^{res} $, so that pure interaction of $N$ Rydberg
atoms can be studied even with poor detectors, provided $\bar {n}$ is sufficiently low. For example, for $N = 2$
and $T = 0.1$ we must use $\bar {n} \ll 1$. In this case the necessary accumulation time is rather long, since the
mean number of atoms detected per laser pulse is small ($\bar {n}T \ll 0.1$). For a good detector with $T = 0.9$
we estimate $\bar {n} \ll 10$ and the accumulation time is two orders of magnitude shorter. From this point of
view, the tolerable detection efficiency is estimated as $T \ge 0.5$, while the appropriate mean number of Rydberg
atoms is $\bar {n} < 1$.

Other interesting observations can be made for the relationships between various multiatom signals. In particular,
the signals $S_{1} $ and $S_{2} $ are of major importance for the observations of the two-atom interactions and
dipole blockade effect. For the Poisson distribution the ratio of their resonant parts is

\begin{equation}
\label{eq29} \alpha = \frac{{S_{1} - \rho _{1}} }{{S_{2} - \rho _{1}} } = \frac{{\bar {n}\left( {1 - T} \right)
\;}}{{1 + \,\bar {n}\left( {1 - T} \right)}}
\end{equation}

We can consider various limits of this relation. It is zero at $T=1$ since in this case there is no resonance in
$S_1$. If $\bar {n}\left( {1 - T} \right) \ll 1$, it is close to $\bar {n}\left( {1 - T} \right)$. In the
intermediate case $\bar {n}\left( {1 - T} \right)\sim 1$ the resonance in $S_{1} $ is about two times smaller than
in $S_{2} $. However, at $\bar {n}\left( {1 - T} \right) \gg 1$ it approaches 1 and the resonances in $S_{1} $ and
$S_{2} $ become of identical amplitudes and shapes, so that it would be impossible to investigate, e.g., pure
two-atom interactions.

We have noticed that a measurement of $\alpha $ provides a straightforward determination of $\bar {n}\left( {1 -
T} \right)$ without any knowledge about $\rho_2^{res}$:

\begin{equation}
\label{eq30} \bar {n}\left( {1 - T} \right) = \frac{{\alpha} }{{1 - \alpha} }
\end{equation}

On the other hand, the mean number of Rydberg atoms detected per laser excitation pulse is $\beta = \bar {n}T$ and
it can also be measured in experiments, e.g., from the histograms in Fig.1(b). Therefore, the two above
measurements provide an unambiguous determination of the unknown experimental parameters $\bar {n}$ and $T$:

\begin{equation}
\label{eq31} \left\{ {{\begin{array}{*{20}l}
 {\bar {n} = \displaystyle\frac{{\alpha} }{{1 -
\alpha} } + \beta}  \\ \\
 {T = \beta /\bar {n}} \\
\end{array}} } \right.
\end{equation}

These formulas are advantageous as they are independent of the specific experimental conditions (atom density,
laser intensity, excitation volume, dipole moments, etc.), which are hard to measure in order to find $\bar {n}$
and $T$ with conventional methods. In fact, Eq.(\ref{eq31}) provides a new method of the absolute $\bar {n}$ and
$T$ measurements. The calculated dependences of \textit{T} and  $\bar {n}$ on the measured parameters $\alpha $
and $\beta $ are show in Fig.3.

\begin{figure}
\includegraphics[scale=0.535]{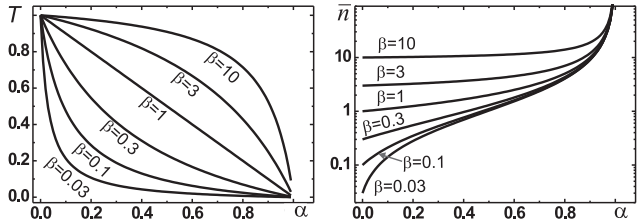}
\caption{\label{Fig3}Calculated dependences of detection efficiency \textit{T} and mean Rydberg atom number $\bar
{n}$ on the parameters $\alpha = \left( {S_{1}^{weak} - \rho _{1}}  \right)/\left( {S_{2}^{weak} - \rho _{1}}
\right)$ and $\beta = \bar {n}T$, which can be measured in experiments.}
\end{figure}

\begin{figure*}
\includegraphics[scale=0.8]{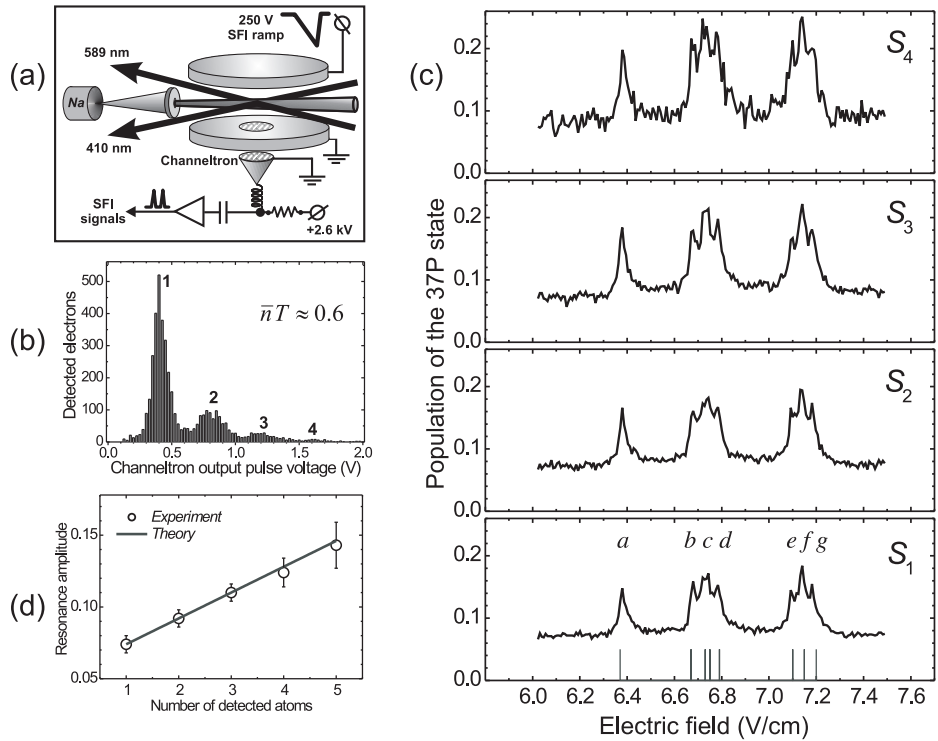}
\caption{\label{Fig4} (a) An arrangement of the experiment on resonant collisions in Na atomic beam. (b) The
histogram of the amplified output pulses of the channeltron measured at the mean number of atoms detected per
laser pulse  $\bar {n}T\approx0.6$. (c) The experimental spectrum of collisional resonances observed in the Na
atomic beam for 1-4 detected Rydberg atoms. The vertical lines indicate the calculated positions of resonances for
various fine-structure components of P states. (d) The comparison between experiment and theory for the dependence
of the amplitude of resonance \textit{a} on the number of detected Rydberg atoms. }
\end{figure*}

\section{Experiment with a sodium atomic beam}

The above theoretical analysis of the Rydberg atom statistics at the SFI detection of resonant collisions needs an
experimental confirmation, since no experiments were made earlier for a few Rydberg atoms. We therefore have
performed the first testing experiment with a sodium atomic beam, where post-selected signals of resonant
collisions were studied for a few detected Rydberg atoms.

\subsection{Experimental setup}

An experimental setup was the same as we used in our recent experiments on collisional and BBR ionization of
sodium Rydberg atoms [34]. The Na effusive atomic beam with the 650~K temperature ($v_0\approx 680$ m/s) was
formed in a vacuum chamber with the 5$\times$10$^{-7}$~Torr background pressure. The beam was passed between the
two stainless-steel plates separated by 10~mm [Fig.4(a)]. A ramp of the pulsed voltage was applied to the upper
plate for the SFI detection of Rydberg atoms. The lower plate had a 10-mm-diameter hole covered by a mesh with
70\% geometrical transparency. The electrons that have appeared in the SFI were passed through this mesh to an
input window of a channeltron. A weak dc electric field was added to the upper plate for the Stark tuning of
resonant collisions.

The 37S$_{1/2}$ state was excited via the two-step scheme $3\mathrm{S}_{1/2}\to 3\mathrm{P}_{3/2}\to
37\mathrm{S}_{1/2}$. The first step at 589~nm was driven by either pulsed (5~kHz) or cw Rhodamine 6G dye-laser.
The broadband pulsed radiation ($\simeq$20~GHz linewidth) could excite the atoms of all velocities, while the
narrowband cw radiation ($\leq$10 MHz linewidth) provided a velocity selection at the cycling hyperfine transition
3S$_{1/2}$(F=2)$ \to $3P$_{3/2}$(F$^\prime$=3). The second step at 410 nm was driven by the second harmonic of a
pulsed titanium-sapphire laser having 5~kHz repetition rate, 50~ns pulses, and 10~GHz linewidth.

In order to work with a few strongly interacting Rydberg atoms, we had to localize a small excitation volume. We
attained this using a crossed-beam geometry shown in Fig.4(a). The two exciting laser beams were focused and
intersected at the right angles inside of the atomic beam. Both laser beams were set at the 45$^\circ $ angles
with respect to the atomic beam. In this way we formed the excitation volume of about 50~$\mu $m size that
provided the mean spacing of a few microns between Rydberg atoms. The laser powers were adjusted to detect a few
Rydberg atoms per laser pulse on the average.

The timing diagram of signals was identical to that of Fig.1(a) with the free DD interaction time $t_0=3$~$ \mu$s.
The channeltron output pulses from the 37P and (37S+36P) states were detected using two independent gates. The
channeltron provided the atom-number resolution according to the measured histogram of Fig.4(b). Although the
neighboring multiatom signals in the histograms are overlapping to some extent, the estimated fidelity of the
correct \textit{N} determination is $>$90$\%$ for the single-atom signal and $>$80$\%$ for the two-atom signal.
The signals were automatically post-selected and sorted out according to the number of detected Rydberg atoms.

\subsection{Experimental results}

The experiments on resonant collisions were performed at the volume density $ n_0 \approx 8\times 10^{10}$
cm$^{-3}$ of ground-state atoms. This value corresponds to the average spacing between neighboring atoms $R_{0}
\approx 1.44\;\mu m$ and to the atom number $N_{0} \approx 10^4$ in the $\mathrm{V} \approx 50 \times 50 \times 50
\;\mu $m$^{3}$ excitation volume. At this density the collision resonances were reliably observed in the $S_{1} $
signal with both pulsed and cw lasers at the first excitation step. The narrow resonances obtained with cw laser
were presented in Fig.2(b). Unfortunately, the use of cw laser significantly reduced the number of atoms excited
to the intermediate 3P$_{3/2}$ state due to the velocity selection and partial optical pumping to the
3S$_{1/2}$(F=1) ground state, so that detection of resonances in the $S_{2} - S_{4} $ signals required too long
accumulation time. Therefore, the further experiments were performed with the pulsed laser in order to provide an
appropriate signal-to-noise ratio in the multiatom signals, at the cost of somewhat broader resonances.

The multiatom spectra of resonant collisions $S_{1} - S_{4} $ observed with the pulsed 589 nm laser are presented
in Fig.4(c). The spectrum for $S_{5} $ is not shown, since it had significant noise due to low probability of the
detection of 5 atoms [see the histogram in Fig.4(b)]. The seven peaks $a-g$ in Fig.4(c) are due to various
transitions to the fine-structure components of P states (actually, nine peaks must appear, but peaks $c$ and $f$
have two unresolved components). The positions of the resonances well coincide with the values obtained from the
numerically calculated Stark map of the relevant Rydberg states [the vertical lines in Fig.4(c)]. In agreement
with Ref. [13], the resonances tend to have a cusp shape, since the pointed tops of the two unresolved resonances
in peak $c$ are well seen.

We will consider in more detail the resonance $36\mathrm{P}_{J = 3/2,\,\left| {M_{J}}  \right| = 1/2}
\leftrightarrow 37\mathrm{S} \leftrightarrow 37\mathrm{P}_{J = 3/2,\,\left| {M_{J}} \right| = 1/2} $ that
corresponds to the well-resolved single peak \textit{a} at $6.37$~V/cm. Its FWHM is $(35\pm3)$~mV/cm. This width
can be converted to the effective frequency width using the polarizabilities measured in our earlier microwave
experiments [35]. The FWHM of resonance $a$ is $(53\pm 5)$~MHz. This experimental value well agrees with the
roughly estimated width $1/(2\pi\tau)\sim v_0/(4\pi R_0)\approx 40$~MHz.

As predicted by the theory, the amplitudes of the resonances in Fig.4(c) grow with the increase of the number of
detected Rydberg atoms, although we expected this effect to be more pronounced. In order to compare the observed
amplitude of resonance $a$ with theory, the radial parts of the dipole moments of the 37S$-$36P and 37S$-$37P
transitions have been numerically calculated and found to be 1372 and 1439 a.u., correspondingly. The angular
parts of both dipole moments are $\sqrt {2} /3$. From Eq.(\ref{eq21}) and Eq.(\ref{eq22}) we have obtained $\Omega
_{0} /\left( {2\,\pi} \right) \sim 150$~MHz and $\Omega /\left( {2\,\pi} \right) \sim 1.5$~MHz. Then
Eq.(\ref{eq25}) yields the theoretical value $\rho _{2}^{res} \left( {\Delta = 0} \right) \sim 0.03$.

Further, at the pulsed laser excitation from the histogram of Fig.4(b) we have measured that the integral of the
two-atom peak (taken over 0.6$-$1~V) is nearly 3.3 times smaller than the integral of the single-atom peak (taken
over 0.2$-$0.6~V). Their ratio can be found from Eq.(\ref{eq4}) as $\bar {P}_{2}^{weak}/\bar {P}_{1}^{weak} = \bar
{n}T/2$, hence $\beta = \bar {n}T \approx 0.6$ in our experiment with the pulsed laser. The parameter $\alpha$ of
Eq.(\ref{eq29}) was determined from the observed relation $\left( {S_{1} - \rho _{1}} \right)/\left( {S_{2} - \rho
_{1}}  \right) \approx 0.075/0.095\approx 0.8$ between the single- and two-atom amplitudes of peak \textit{a} in
Fig.4(c). Finally, Eq.(\ref{eq31}) yields the unknown experimental parameters $\bar {n} \approx (4.7\pm 0.5)$ and
$T \approx (13\pm 1.5)\,\% $.

This relatively low detection efficiency was unexpected, since we specially installed a newly made channeltron and
expected \textit{T} to be close to the geometrical transparency of the detector mesh (70\%). Our attempts to
improve the SFI detection system (acceleration of electrons prior to detection, detection of ions instead of
electrons, cleaning of the channeltron's surface, etc.) did not noticeably increase the efficiency.

This observation contrasts with the recent paper [36], where simultaneous detection of ions and electrons,
delivered from photoionization of Rb atoms, by two opposite channeltrons allowed to observe the 50\% detection
efficiency for ions. A dependence of the detection efficiency on the channeltron's voltage has been additionally
investigated in this work. It exhibited a strongly nonlinear behavior, with zero efficiency at 2.4 kV and 50\%
efficiency at 2.9 kV. Unfortunately, in our experiment we applied the 2.6 kV voltage specified by the manufacturer
and did not vary it in a broad range. This seems to be the most likely reason for the lower detection efficiency
in our case. This surmise will be tested in a forthcoming experiment with cold Rb Rydberg atoms. Another way to
improve the detection efficiency could be the usage of the SFI detection system based on a set of resistively
coupled parallel cylindrical plates, which provides a homogeneous electric field without a mesh [37].

Nevertheless, we have demonstrated that the above measurements provide a tool to determine unknown \textit{T},
which may degrade during the course of any experiments and therefore should be periodically checked out. We may
recommend our method to be applied in those experiments with Rydberg atoms where the absolute detection efficiency
is crucial.

Finally, with the above experimental values for $\bar {n}$, $T$, and amplitude $S_{1} - \rho _{1} \approx 0.075$
of peak \textit{a}, using Eq.(\ref{eq27}) we have determined the experimental value $\rho _{2}^{res} \left(
{\Delta = 0} \right) \approx (0.02\pm 0.004)$. It satisfactory agrees with the theoretical value 0.03, taking into
account the approximations adopted in our theoretical model. Furthermore, we can compare the values for the
amplitudes of observed resonances to those calculated with Eq.(\ref{eq27}) at $\rho _{2}^{res} \left( {\Delta = 0}
\right) = 0.02$. These are shown in Fig.4(d), where we also added an experimental point for $S_{5}$ extracted from
its noisy spectrum. The experimental and theoretical dependences coincide well for all \textit{N}, showing a
linear dependence and confirming the validity of Eq.(\ref{eq27}). A comparison between experiment and theory for
the other peaks in Fig.4(c) has also demonstrated a satisfactory quantitative agreement.

\section{Conclusions}

The developed theoretical model describes multiatom signals that are measured in the experiments on long-range
interactions of a few Rydberg atoms. Although several simplifying approximations were used to obtain the
analytical formulas, their validity has been partly confirmed by the satisfactory agreement with the first
experimental data obtained for the resonant collisions of a few Rydberg atoms in the atomic beam. The main result
is that finite detection efficiency of the selective field-ionization detector leads to the mixing up of the
spectra of the resonant collisions associated with various numbers of actually excited atoms. In particular, the
collisional resonance features may appear even in the single-atom signal if the detection efficiency is low. This
may misinterpret the possible observations of the full dipole blockade or coherent two-atom collisions, which are
required for quantum logic gates. The obtained formulas are helpful in estimating the appropriate mean number of
Rydberg atoms excited per laser pulse at a given detection efficiency.

We have also shown that a measurement of the relation between the amplitudes of resonances observed in the single-
and two-atom signals provides a straightforward determination of the absolute detection efficiency and mean
Rydberg atom number. This new method is advantageous as it is independent of the specific experimental conditions
(atom density, laser intensity, excitation volume, dipole moments, etc.).

The further experimental efforts should be focused on cold Rydberg atoms. This would help to test our model at the
longer interaction times, which are of interest to quantum information processing. Microwave spectroscopy of
resonant collisions [38,~39] can be additionally applied as a sensitive probe and control tool.

\begin{acknowledgments}
The authors are indebted to G.~I.~Surdutovich, A.~M.~Shalagin, and A.~Stibor for fruitful discussions. This work
was supported by the Russian Foundation for Basic Research, Grant No. 05-02-16181, by the Russian Academy of
Sciences, and by INTAS, Grant No. 04-83-3692.
\end{acknowledgments}

\end{document}